\documentclass[aps,nofootinbib,floats,superscriptaddress]{revtex4}

\usepackage{graphicx}
\usepackage{epstopdf}
\usepackage{amssymb}
\usepackage{amsmath,bm}
\usepackage{psfrag}
\usepackage{epsfig}
\usepackage{float}
\usepackage{bm}
\newcommand{\angstrom}{\mbox{\normalfont\AA}}

\begin{document}

\title{Kinetic models of Quantum Size Effect-directed nanocluster self-assembly in atomic corrals}

\author{Mikhail Khenner\footnote{Corresponding author. E-mail: mikhail.khenner@wku.edu.}}
\affiliation{Department of Mathematics, Western Kentucky University, Bowling Green, KY 42101, USA}
\affiliation{Applied Physics Institute, Western Kentucky University, Bowling Green, KY 42101, USA}

\begin{abstract}
\noindent
Two simple kinetic models of Quantum Size Effect-directed nanocluster self-assembly in circular atomic corrals are discussed. 
The models correspond to an adsorption (either a physisorption or a chemisorption) and an adsorption-diffusion 
regimes that are typical at low and high temperatures, respectively. Small magnitudes of a variation of the electronic 
local density of states is shown to be the prime factor that impedes self-assembly in the latter regime.

\end{abstract}

\date{\today}
\maketitle


\section{Introduction}
\label{Intro}

The scattering of the surface electrons off surface-implanted atoms generates Friedel oscillations on the close-packed surfaces of
noble metals, i.e. a standing wave patterns in the electron
density (Local Density of States, or LDOS) \cite{CrommieEtAl}. A standing wave pattern is the result of interference of two or more scattered waves. 
Standing wave patterns can be used to confine adatoms inside quantum corrals built from surface-implanted atoms and thus engineer 
artificial structures of single adatoms or self-assembled nanoclusters.   

Despite that standing wave LDOS patterns on surfaces and a long-range adsorbate interactions mediated by a two-dimensional electron gas 
are relatively well-understood \cite{CrommieEtAl,KBESHK,RMMRHP,HP}, the kinetics of self-assembly driven by standing wave LDOS patterns 
was not sufficiently studied neither experimentally, nor theoretically. For theoretical support, sometimes an experimental 
study of self-assembly outcomes is supplemented with a limited Monte-Carlo simulation of adatom diffusion \cite{CMZSYZWHBD,LCD,SNNLB}, 
but basic questions 
such as to what extent the geometry of a standing wave LDOS pattern and a magnitude of a spatial variation of LDOS affect 
self-assembly and its kinetics are left unclarified.
In fact, LDOS is seldom constructed and directly appealed to when interpreting experimental outcomes.
In this notice, we construct two minimalist models of self-assembly kinetics in order to point out that these factors are crucial.

QSE diminishes with rising temperature, and typically at room temperature it has a vanishingly small effect on self-assembly. However, in some systems 
\cite{MRDS,KSRCLS,KWHDS,JMNSCJX,GCWYTLL} QSE drives the self-assembly even at room temperature. It was argued that in these systems 
QSE selectively affects the adsorption properties of the substrate, i.e. it laterally modulates the adsorption, thus the atoms that are deposited onto 
a strong-adsorption sites bind strongly to the substrate \cite{MRDS,JMNSCJX}.  As clusters of several atoms form, the electronic shell effect of 
quantum confined electrons may further increase binding of atoms in a cluster and even promote nucleation of “magic” clusters \cite{GCWYTLL}.
At room temperature the surface diffusion is not sufficiently strong  to dissolve the newly formed clusters.
Correspondingly, our models describe a low (high) temperature regime when the kinetics is governed by adsorption (by adsorption and diffusion). 
We take all required QSE-related material parameters directly from Ref. \cite{MRDS}, where they correspond to Ni adsorbate on Rh(111) substrate. 
This is for illustration only, as the models can be applied to any adsorbate/substrate pair that produces QSE-directed self-assembly via 
spatial modulation of adsorption.

Ref. \cite{MRDS} posits that a standing wave pattern of LDOS modifies the adsorption properties of
the surface. 
The variation of the LDOS ($\Delta$LDOS) depends on Fermi wavenumber $k_F$, the distance $r_i$ from the
$i$-th scattering center and the Fermi level phase shift $\delta$, 
 and is described by the equation \cite{CrommieEtAl}:

\begin{eqnarray}
\Delta \mbox{LDOS} &=& \sum_{i=1}^n \Delta \mbox{LDOS}_i = \sum_{i=1}^n \frac{\cos^2{\left(2\pi r_i -\pi/4  +\delta\right)} - 
\cos^2{\left(2\pi r_i - \pi/4\right)}}{2\pi r_i}, \label{DeltaLDOS}\\
r_i &=& \sqrt{\left(x - x_c^{(i)}\right)^2 + \left(y - y_c^{(i)}\right)^2}. \nonumber
\end{eqnarray}

Eq. (\ref{DeltaLDOS}) states that $\Delta \mbox{LDOS}$ is the sum of contributions from $n$ scattering centers (metal atoms) located at 
$\left(x_c^{(i)},y_c^{(i)}\right)$. A maximum in $\Delta \mbox{LDOS}$ is a potential well and it is a
suitable adsorption site. In this paper, the scattering centers are assumed to be equispaced on a circle of radius $R$ (a corral), thus
\begin{equation}
x_c^{(i)} = R \cos{\frac{2\pi i}{n}},\quad y_c^{(i)} = R \sin{\frac{2\pi i}{n}},
\quad i=1,...,n. \label{scatters}
\end{equation}
All distances were made non-dimensional with Fermi wavelength $\lambda_F$.
$\delta=0.3$ \cite{MRDS} will be used for all $\Delta$LDOS computations in this paper.

\begin{figure}[H]
\vspace{-0.2cm}
\centering
\includegraphics[width=4in]{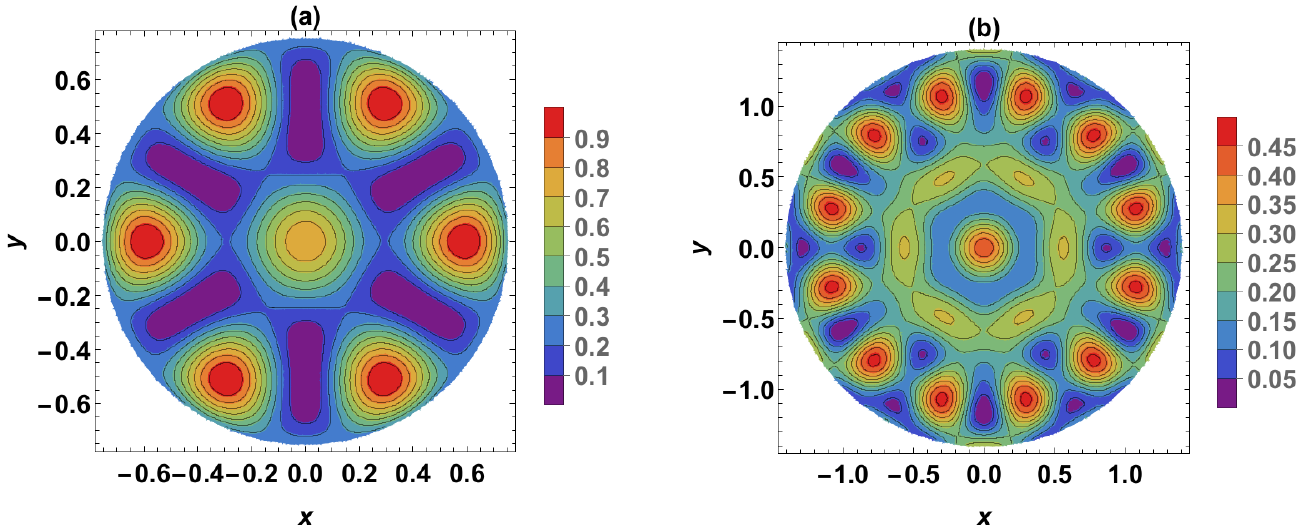}
\vspace{-0.15cm}
\caption{$\Delta$LDOS in the interior of a six-atoms corral. The atoms (scatterers) themselves, and $\Delta$LDOS in their immediate vicinity are not shown. 
(a) $R=1$. (b) $R=2$. Note from the color bars that $M_\Delta\approx 1$ in (a), $M_\Delta\approx 0.5$ in (b). For comparison with (b), 
$\Delta$LDOS in the interior of a four, eight and twelve-atoms corrals of the same radius is shown in Fig. \ref{Fig6}.
}
\label{Fig1}
\end{figure}

Figures \ref{Fig1}(a,b) show $\Delta$LDOS in the interior of two corrals of different radii, prepared with $n=6$ scattering centers 
with the angular separation of 60$^\circ$.\footnote{We consider small corrals (both size-wise and atoms-wise) in order to compute $\Delta \mbox{LDOS}$ 
and adsorbate coverage with high accuracy.}  Notice that the magnitude of $\Delta \mbox{LDOS}$, i.e. the difference between $\Delta \mbox{LDOS}$ maximum and minimum, 
is two times larger for $R=1$ case. In this paper we denote the magnitude of $\Delta \mbox{LDOS}$ as $M_\Delta$. The expression for $M_\Delta$ can't be easily determined 
from Eq. (\ref{DeltaLDOS}) even for simple scattering geometries such as corrals made of a handful of atoms. Also,  
there is no easy or convenient way to analytically find or estimate the number and location of peaks in $\Delta$LDOS for 
given $\delta$, $n$ and $R$. Thus the contour plots such as in 
Fig. \ref{Fig1} are the most practical way to quantify $M_\Delta$ and to determine the landscape of $\Delta \mbox{LDOS}$.
\footnote{From Eq. (\ref{DeltaLDOS}) it is obvious that $\Delta$LDOS takes on negative values at some points $(x,y)$. 
For uniformity, we shift all computed $\Delta$LDOS into a positive range by a pointwise addition of a constant 
$\mbox{abs}\left[\mbox{min}\left(\Delta\mbox{LDOS}\right)\right]$, where the minimum is computed over all points inside a disk (see Figures \ref{Fig1} and \ref{Fig6}). 
Such shifts by a constant do not affect any results, but they simplify the comparisons of various $\Delta\mbox{LDOS}$.}

\section{Adsorption Modeling (A-model)}
\label{A}

We start by assuming that at low temperatures ($< 300$ K) a surface diffusion is negligible in comparison to adsorption.
Thus in this section we apply $\Delta$LDOS shown in Figures \ref{Fig1}(a,b) to compare the adsorption kinetics inside the corresponding corrals.

Although atomic and molecular adsorption is a complicated and heavily material-dependent phenomenon on a microscopic level \cite{LWHLTYH}, the 
adsorption on a coarse, i.e. mesoscopic, level is typically differentiated into a physisorption and a chemisorption \cite{Luth}. 
The prototype kinetics of physisorption is usually modeled by an initial value problem for a first-order ordinary differential equation (ODE IVP) and is known as 
Pseudo-First Order (PFO) kinetics \cite{Azizian}. 
In the context of continuous adsorption that is mediated by $\Delta \mbox{LDOS}$ this problem reads:
\begin{equation}
\frac{d\rho}{dt} = A\; \Delta \mbox{LDOS}\; (1 - \rho), \quad \rho(x,y,0)=\rho_0(x,y). \label{adsPFO-Eq-final}
\end{equation}
Here $\rho(x,y,t)$ is adsorbate coverage in monolayers ($0\le \rho\le 1$), $\rho_0(x,y)$ is the initial coverage, and $A$ is the dimensionless parameter that quantifies the adsorption strength.
Analytical solution of the problem (\ref{adsPFO-Eq-final}) is easily found:
\begin{equation}
\rho_{\mbox{pfo}} = 1-\exp{\left(-A\; \Delta \mbox{LDOS}\; t\right)} \left(1-\rho_0\right).
\label{adsPFO-Sol-final}
\end{equation}


Similarly, the prototype ODE IVP for chemisorption kinetics is known as Pseudo-Second Order (PSO) kinetics \cite{Azizian}. For adsorption that is modulated by 
$\Delta \mbox{LDOS}$ that problem reads:
\begin{equation}
\frac{d\rho}{dt} = A\; \Delta \mbox{LDOS}\; (1 - \rho)^2, \quad \rho(x,y,0)=\rho_0(x,y). \label{adsPSO-Eq-final}
\end{equation}
Its analytical solution is:
\begin{equation}
\rho_{\mbox{pso}} = \frac{\rho_0+A\; \Delta \mbox{LDOS}\;\left(1-\rho_0\right)t}{1+A\; \Delta \mbox{LDOS}\;\left(1-\rho_0\right)t}. \label{adsPSO-Sol-final}
\end{equation}

\begin{figure}[H]
\vspace{-0.2cm}
\centering
\includegraphics[width=4.5in]{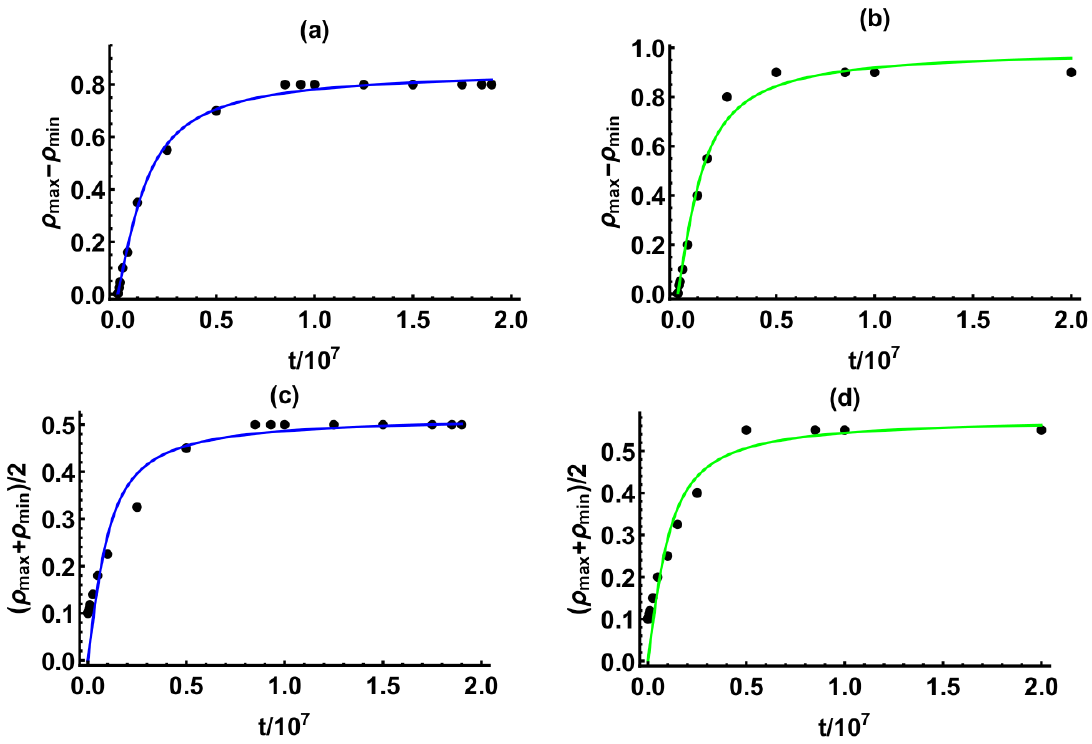}
\vspace{-0.15cm}
\caption{
A-model: Comparison of adsorption kinetics for the six-atoms corral with $\Delta \mbox{LDOS}$ in Fig. \ref{Fig1}(b).
(a) $\rho_{max}-\rho_{min}$, PSO. Fit: $0.547 \arctan{(6.903 t)}$. 
(b) $\rho_{max}-\rho_{min}$, PFO. Fit: $0.634 \arctan{(8.146 t)}$.   
(c) Median $\rho$, PSO. Fit: $0.330 \arctan{(10.272 t)}$. 
(d)  Median $\rho$, PFO. Fit: $0.369 \arctan{(9.969 t)}$.   
}
\label{Fig3a}
\end{figure}

Note that the time $t$ is dimensionless in Eqs. (\ref{adsPFO-Eq-final})-(\ref{adsPSO-Sol-final}). 
For consistency with 
the adsorption-diffusion model (A+D model, Sec. \ref{AD}), the physical time 
was adimensionalized by the same scale as in Sec. \ref{AD}, i.e. $\bar t=\lambda_F^2/\left(\nu^2 \mathcal{D} S^2\right)$, 
where $\mathcal{D}$ is the surface diffusion coefficient of an adsorbate, $\nu$ the aerial
density of adsorption sites, and $S=\pi d^2/4$ the area of an adsorbate 
atom cross-section (with $d$ the diameter of an adsorbate atom).  
$\bar t = 4\times 10^{-6}$ s at the typical values $\lambda_F=2\pi/k_F=0.9$ nm (corresponding to $k_F=0.7\;\angstrom^{-1}$ \cite{MRDS}), $\nu=5\times 10^{14}$ cm$^{-2}$, 
$\mathcal{D}=10^{-8}$ cm$^2$/s (at $T=300$ K), $S=8.35\times 10^{-16}$ cm$^2$ 
(corresponding to $d=0.326$ nm of Ni atom). 
 Thus $A$ was chosen of similar magnitude, $A=10^{-6}$, and the solutions 
$\rho_{\mbox{pfo}}$ and $\rho_{\mbox{pso}}$ were evaluated to $t\sim 2\times 10^7$, which corresponds to a physical time about 80 s (see Figure \ref{Fig3a}). 

From Eqs. (\ref{adsPFO-Sol-final}) and (\ref{adsPSO-Sol-final}) it is easy to notice that for any positive $A$ and $\Delta \mbox{LDOS}$, 
$\rho_{\mbox{pfo}}$ and $\rho_{\mbox{pso}}$ approach one (the complete monolayer coverage) as $t$ tends to infinity, albeit the kinetic laws are different 
and the exact kinetics should be sensitive to $M_\Delta$ and, to a lesser degree, to (spatial) variations of $\Delta \mbox{LDOS}$. In Fig. \ref{Fig3a} these differences
are quantified. Compared to PSO, PFO adsorption kinetics establishes the maximum spatial difference of the coverage and the median value of the coverage 
over the substrate area inside a corral more than twice faster. Because of faster kinetics, the final values of these quantities at $t=2\times 10^7$ for PFO case 
are 15\% larger than the corresponding values for PSO case.     

\begin{figure}[H]
\vspace{-0.2cm}
\centering
\includegraphics[width=6.5in]{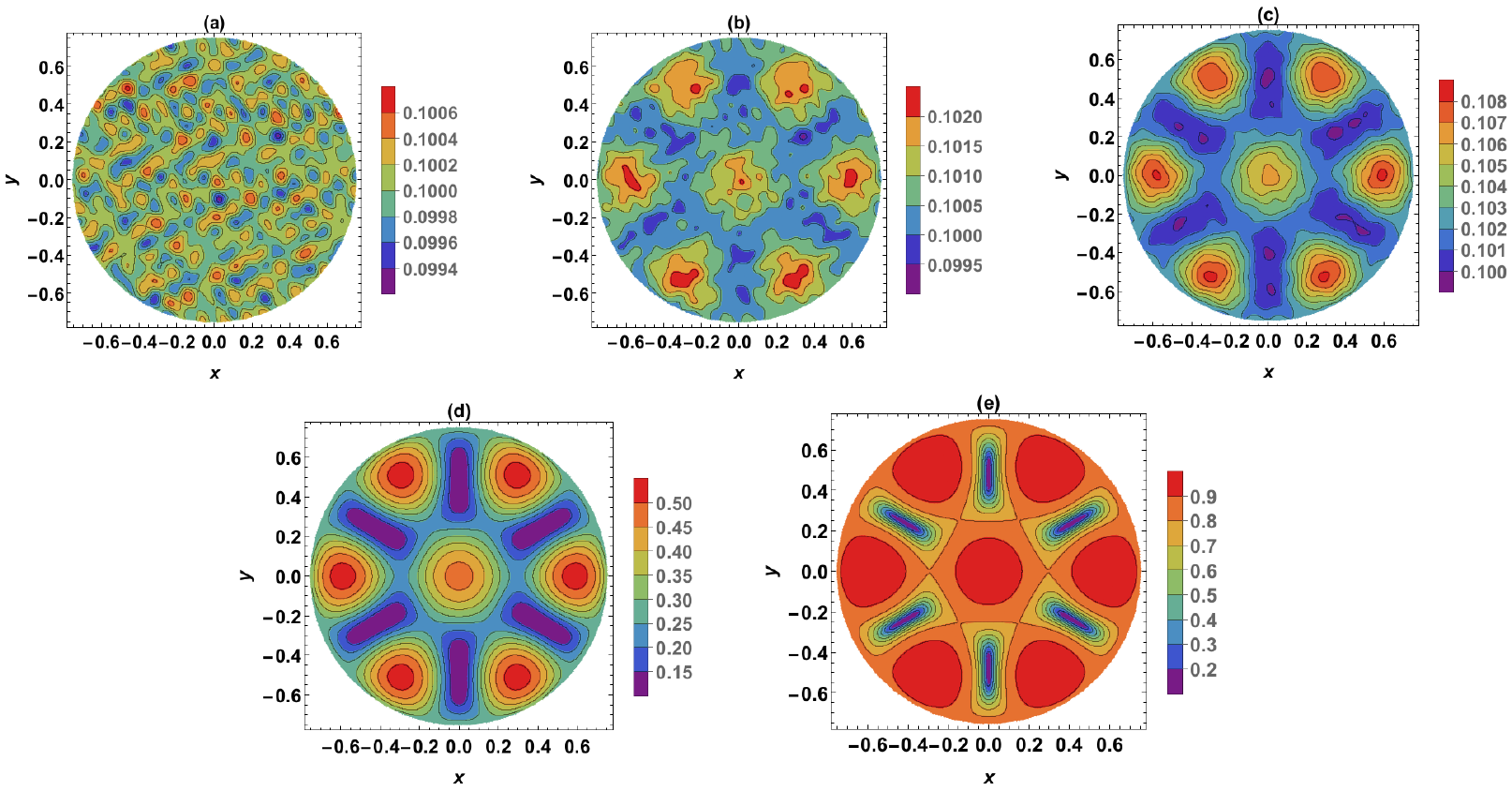}
\vspace{-0.15cm}
\caption{A-model: $\rho_{\mbox{pso}}$ at the increasing times for the six-atoms corral with $\Delta \mbox{LDOS}$ in Fig. \ref{Fig1}(a). (a) $t=0$, (b) $t=2500$, (c) $t=10^4$, 
(d) $t=10^6$, (e) $t=2\times 10^7$.
}
\label{Fig2}
\end{figure}

\begin{figure}[H]
\vspace{-0.2cm}
\centering
\includegraphics[width=6.5in]{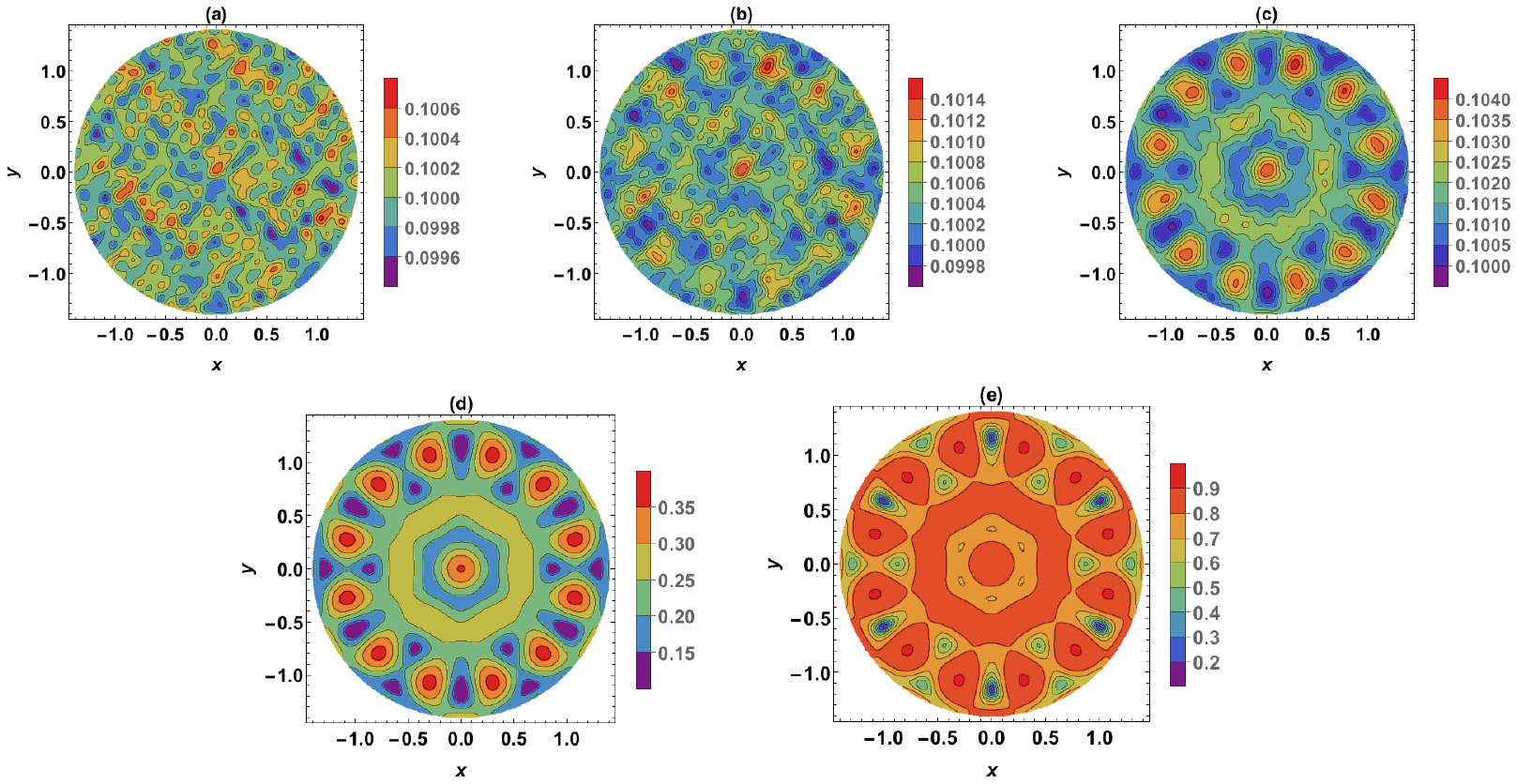}
\vspace{-0.15cm}
\caption{A-model: $\rho_{\mbox{pso}}$ at the increasing times for the six-atoms corral with $\Delta \mbox{LDOS}$ in Fig. \ref{Fig1}(b). (a) $t=0$, (b) $t=2500$, (c) $t=10^4$, 
(d) $t=10^6$, (e) $t=2\times 10^7$.
}
\label{Fig2a}
\end{figure}

The snapshots of $\rho_{\mbox{pso}}$ evolution are shown in Figures \ref{Fig2} and \ref{Fig2a}; we omit similar looking Figures for $\rho_{\mbox{pfo}}$. 
It is seen that the atoms preferentially adsorb at the areas of the substrate where $\Delta \mbox{LDOS}$ is maximized in Figures \ref{Fig1}(a,b). 
These regions are associated with a self-assembled nanoclusters, whose size increases with time. 
At $t=2\times 10^7$ the nanocluster formation is complete ($\rho_{\mbox{pso}}\sim 1$, i.e. a monolayer coverage is locally reached).

\section{Adsorption and Diffusion Modeling (A+D-model)}
\label{AD}

At temperatures above room temperature both adsorption \emph{and} surface diffusion are operative. 
The coverage evolves according to the following dimensionless partial differential equation:
\begin{eqnarray}
\frac{\partial \rho}{\partial t} &=& A\; \Delta \mbox{LDOS}\; (1 - \rho)^k + \bm{\nabla}^2 \rho +\bm{\nabla}\cdot \bm{J},\quad \rho(x,y,0)=\rho_0(x,y), \quad k =1,2\label{rho-eq-final} \\
\bm{J} &=& \left(\nu S\right)^{-1} B \rho\left(1-\rho\right)\bm{\nabla}\rho. \nonumber
\end{eqnarray}
The diffusion parts of Eq. (\ref{rho-eq-final}), i.e. the second and the third terms at the right-hand side, 
were derived from a law of mass conservation \cite{MyJAP}. 
The exponent $k=1$ ($k=2$) in the adsorption term stands for PFO (PSO) adsorption kinetics, $\bm{J}$ is the component of the diffusion flux that emerges 
due to lateral adsorbate-adsorbate interactions, and
\begin{equation}
|B|=\left|\frac{\beta_1+\beta_2}{k T}\right| \label{F1}
\end{equation}
is the dimensionless magnitude of these 
interactions. Eq. (\ref{rho-eq-final}) necessitates a numerical solution.


The first term in the numerator of $B$ is the first moment of the Friedel interaction potential $U_F$ of two adsorbate atoms separated by a 
distance $r$ \cite{MRDS,KBESHK,RMMRHP,HP} (recall, that all distances are made dimensionless using $\lambda_F$ as the length scale):
\begin{eqnarray}
\beta_1&=&-\int_a^\infty U_F(r)\; dr = -E_F \left(\frac{2\sin{\delta}}{\pi}\right)^2 \int_a^\infty \frac{\sin{2(2\pi r+\delta)}}{(2\pi r)^2}\; dr \nonumber \\
&=& \frac{-2E_F \sin^2{\delta}}{\pi^3}\left[2\sin{(2\delta)}\;\mbox{Si}\left(4\pi a\right)-2\cos{(2\delta)}\;\mbox{Ci}\left(4\pi a\right)-\pi \sin{2\delta}+  \frac{\sin{2(\delta+2\pi a)}}{2\pi a}
 \right]. \label{beta1}
\end{eqnarray}
Here $a$ is the (dimensionless) lattice spacing of the adsorbate, 
$\mbox{Si}(z)=-\int_z^\infty \sin{(t)}/t\; dt$ and $\mbox{Ci}(z)=-\int_z^\infty \cos{(t)}/t\; dt$ are the sine and cosine integral functions.
For Ni 
$a=0.277$ and $E_F=5.01$ eV \cite{EPO}.

The second term in the numerator of $B$ is the first moment of the attractive-repulsive potential for the atoms of an adsorbate \cite{SuttonChen}:
\begin{eqnarray}
\beta_2&=&-\int_a^\infty U_{SC}(r)\; dr = -\epsilon\int_a^\infty \left[\left(\frac{a}{r}\right)^n-2c\left(\frac{a}{r}\right)^{m/2}\right]\; dr \nonumber \\
&=&\epsilon a \left(\frac{1}{n - 1} + \frac{2 c}{1 - m/2}\right). \label{beta2}
\end{eqnarray}
Here $\epsilon$ is the magnitude of $U_{SC}(r)$ with the dimension of energy, and $c$, $m$ and $n$ are the dimensionless parameters of the potential.
At $n>m$, $U_{SC}(r)$ is attractive at long distances and repulsive at very short distances, preventing the unphysical collision of the atoms.
For Ni 
$\epsilon=0.016$ eV, $c=39.426$,  $m=7$, and $n=9$ \cite{SuttonChen}. As evaluated, $\beta_1,\; \beta_2<0$ and they have similar magnitudes, 
$|\beta_1| \sim |\beta_2| \sim 0.006$ eV.

\begin{figure}[H]
\vspace{-0.2cm}
\centering
\includegraphics[width=7.0in]{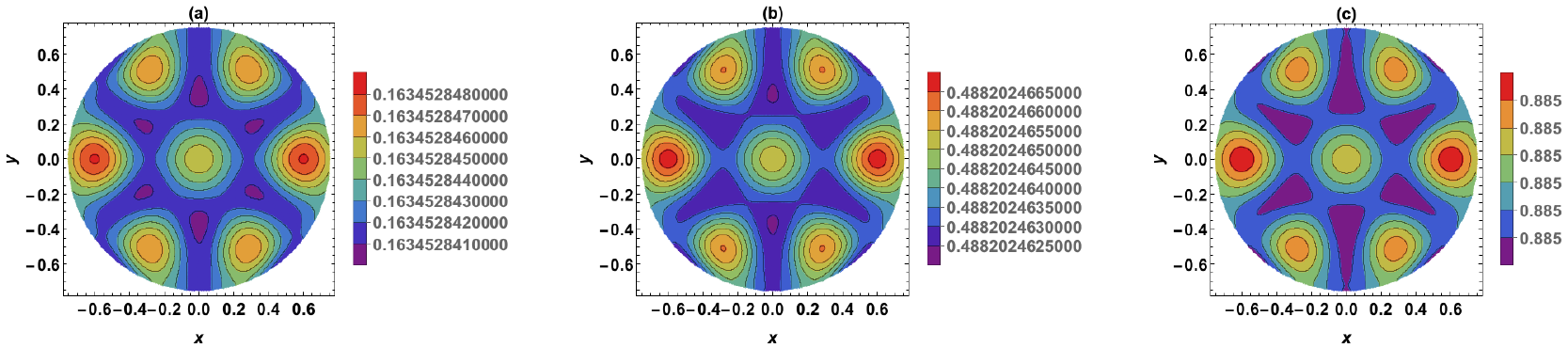}
\vspace{-0.15cm}
\caption{$A+D$ model: $\rho_{\mbox{pso}}$ at the increasing times for the six-atoms corral with $\Delta \mbox{LDOS}$ in Fig. \ref{Fig1}(a). 
(a) $t=10^5$, (b) $t=2\times 10^6$, (c) $t=1.5\times 10^7$.
}
\label{Fig4}
\end{figure}

\begin{figure}[H]
\vspace{-0.2cm}
\centering
\includegraphics[width=7.0in]{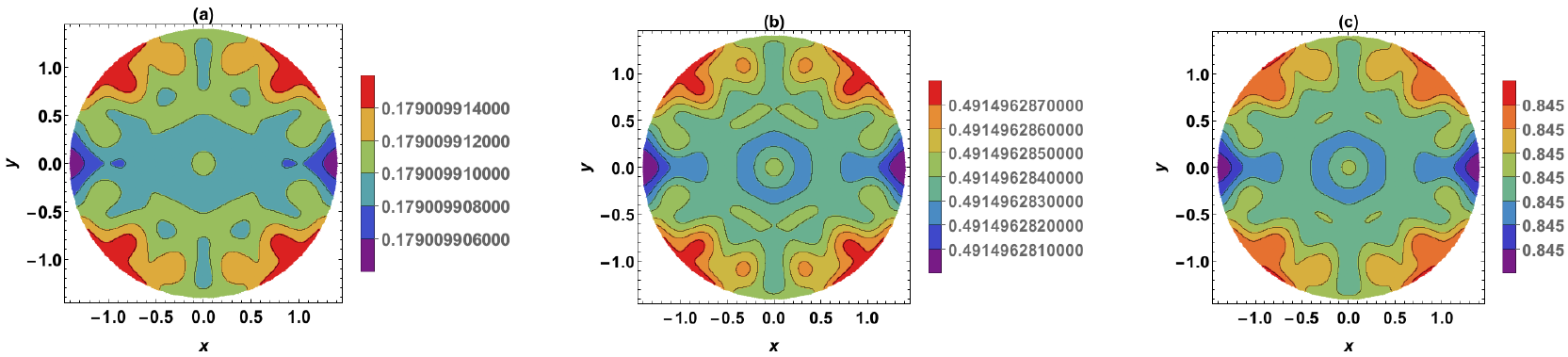}
\vspace{-0.15cm}
\caption{$A+D$ model: $\rho_{\mbox{pso}}$ at the increasing times for the six-atoms corral with $\Delta \mbox{LDOS}$ in Fig. \ref{Fig1}(b). 
(a) $t=10^5$, (b) $t=7\times 10^5$, (c) $t=2.5\times 10^7$.
}
\label{Fig5}
\end{figure}

To illustrate how the surface diffusion 
interferes with $\Delta \mbox{LDOS}$-mediated adsorption
inside the corrals, Figures \ref{Fig4} and \ref{Fig5} show
the evolution of $\rho_{\mbox{pso}}$ computed using Eq. (\ref{rho-eq-final}) with $\Delta \mbox{LDOS}$ in Figures \ref{Fig1}(a,b), 
i.e. for the small and large six-atoms corrals.  
These results should be compared to
Figures \ref{Fig2} and \ref{Fig2a} that show a nanocluster self-assembly inside these corrals due only to $\Delta \mbox{LDOS}$-mediated adsorption. 
(The initial condition $\rho(x,y,0)=\rho_0(x,y)$ is the same for the computations of A-model and A$+$D model.) 
It is seen that the median coverage still increases due to continuous adsorption but diffusion, as expected, makes $\rho_{max}-\rho_{min}$ 
close to negligible. Note however, that the footprint of $\Delta \mbox{LDOS}$-mediated adsorption remains in Fig. \ref{Fig4}, but it is gone in Fig. \ref{Fig5}. 
Microscopically, a nanoclusters comprised of a few atoms are expected to form at the locations of $\Delta \mbox{LDOS}$ maxima inside the smaller corral of radius one, but they 
are not expected to form inside the larger corral of radius two. This crucial difference can be attributed only to larger $M_\Delta$ for $\Delta \mbox{LDOS}$ in Fig. \ref{Fig1}(a).


\begin{figure}[H]
\vspace{-0.2cm}
\centering
\includegraphics[width=7.0in]{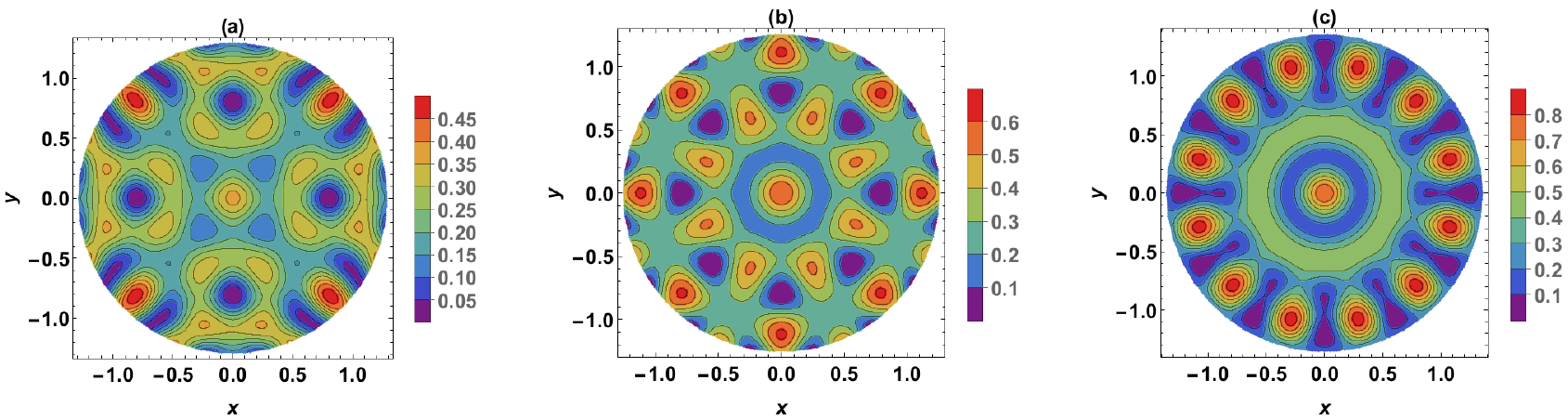}
\vspace{-0.15cm}
\caption{$\Delta$LDOS in the interior of a corral with $R=2$ and 
(a) four scattering centers, (b) eight scattering centers, (c) twelve scattering centers equispaced on a corral boundary.
The scattering centers themselves and $\Delta$LDOS in their immediate vicinity are not shown.
}
\label{Fig6}
\end{figure}

\begin{figure}[H]
\vspace{-0.2cm}
\centering
\includegraphics[width=7.0in]{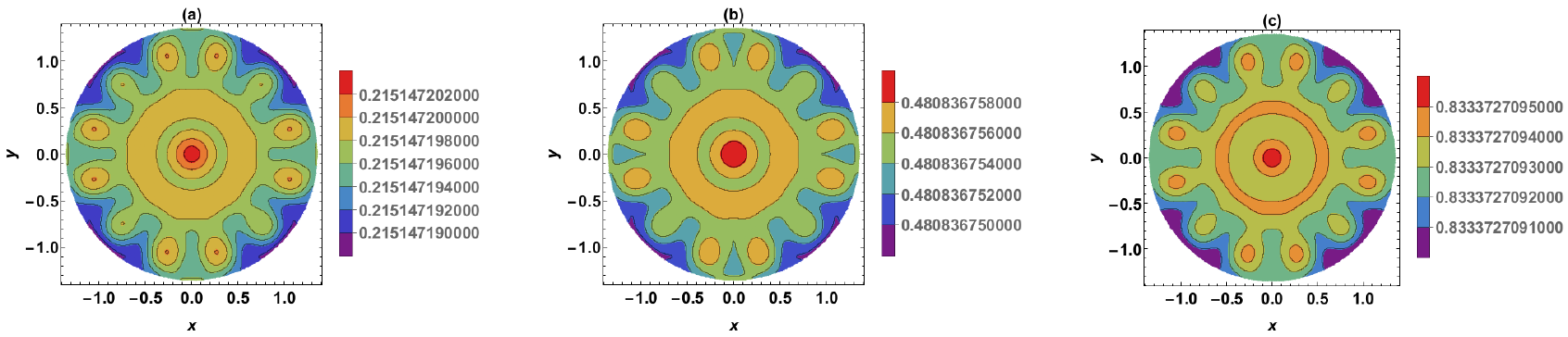}
\vspace{-0.15cm}
\caption{$A+D$ model: $\rho_{\mbox{pso}}$ at the increasing times for the twelve-atoms corral with $\Delta \mbox{LDOS}$ in Fig. \ref{Fig6}(c). 
(a) $t=5\times 10^5$, (b) $t=2.5\times 10^6$, (c) $t=1.5\times 10^7$.
}
\label{Fig7}
\end{figure}

To further support this conclusion, in Fig. \ref{Fig6} we show $\Delta \mbox{LDOS}$ in the interior of three circular corrals of radius two. The 
corrals are prepared with four, eight, and twelve scattering centers that are equispaced on the corral's boundary 
($\pi/2$, $\pi/4$, and $\pi/6$ angular separation between the centers). 
$M_\Delta$ increases from $\approx 0.5$ to $\approx 0.9$ as the number of scattering centers increases. Note that 
for four and six scattering centers $M_\Delta$ values are very close, see Fig. \ref{Fig1}(b). It follows that $M_\Delta$ is a monotonically increasing function of the 
number of scattering centers (at least for the corral of radius two). We computed evolution of $\rho_{\mbox{pso}}$ inside these corrals using $A+D$ model. 
For the eight and twelve-atoms corrals that produce larger $M_\Delta$ in $\Delta \mbox{LDOS}$, the coverage evolution is qualitatively similar to one shown 
in Fig. \ref{Fig4}; for the twelve-atoms corral the evolution is shown in Fig. \ref{Fig7}. Nanocluster self-assembly inside these corrals is therefore expected.
For the four-atoms corral that produces small $M_\Delta$ in $\Delta \mbox{LDOS}$, the coverage evolution is qualitatively similar to one shown in
in Fig. \ref{Fig5}. Nanocluster self-assembly inside this corral is not expected. Thus it follows that there exists a threshold value of $M_\Delta$ around 0.7 
below which $\Delta \mbox{LDOS}$-mediated adsorption loses a competition with diffusion and thus it cannot sustain a nanocluster formation inside the corrals.

%

\setcounter{equation}{0}
\section{Conclusions}
\label{Conc}

We formulated the adsorption and the adsorption-diffusion kinetic models for 
the QSE-directed
nanocluster self-assembly 
inside a circular corrals built up of metal atoms that act as scattering centers for the surface electrons. 
In our minimalist modeling, once $\Delta \mbox{LDOS}$ is fixed by
selecting an adsorbate, a substrate, a corral radius and the number of atoms on a corral boundary, the self-assembly kinetics is governed by only one 
dimensionless parameter (the adsorption model) or two dimensionless parameters (the adsorption-diffusion model). 
These parameters are the adsorption rate and the strength of lateral adsorbate-adsorbate interactions.    
The models show that when surface diffusion is significant, as is the case when temperature is high, a self-assembly is sustained only inside 
those corrals that exhibit an above-a-threshold value of $\Delta \mbox{LDOS}$ magnitude. Computations such as performed in this paper may be a 
useful tool for designing corrals that guide a nanocluster self-assembly via the electronic quantum size effects. 

The described $A+D$ model may be tuned and made more quantitative, for instance, by ascribing a temperature dependence to the adsorption strength $A$.
Rather than assuming such dependence based on general considerations, it would be more useful to extract it from the experiment data for a specific 
adsorbate/substrate system. (Likewise, a surface diffusion
coefficient $\mathcal{D}$ should be taken in the Arrhenius form, where the activation energy and the pre-factor are material system-specific.) 
Unfortunately, we were unable to find an account of an experiment done at different temperatures. 
Thus a detailed investigation is postponed until such data becomes available. 



\end{document}